# Bound state in the continuum induced room-temperature superfluorescence

Haijun Tang[1,†], Hamdi Barkaoui[1,†], Can Huang[1], Xiong Jiang[1], Yixuan Zeng[2], Shumin Xiao[1,2,3,4,5], Shaohua Yu[2], Jiecai Han[3], Qinghai Song[1,2,4,5,*]

1. Ministry of Industry and Information Technology Key Lab of Micro-Nano Optoelectronic Information System, Guangdong Provincial Key Laboratory of Semiconductor Optoelectronic Materials and Intelligent Photonic Systems, Harbin Institute of Technology, Shenzhen 518055, P. R. China.

2. Pengcheng Laboratory, Shenzhen 518055, P. R. China.

3. National Key Laboratory of Science and Technology on Advanced Composites in Special Environments, Harbin Institute of Technology, Harbin 150080, P. R. China.

4. Quantum Science Center of Guangdong-Hong Kong-Macao Greater Bay Area, Shenzhen, P. R. China

5. Collaborative Innovation Center of Extreme Optics, Shanxi University, Taiyuan 030006, Shanxi, P. R. China.

† These authors contribute equally to this work.

Corresponding to: *qinghai.song@hit.edu.cn

## Abstract:

**Superfluorescence is a collective emission from several quantum emitters that initially have random phases and are then synchronized through vacuum field interactions.[1] Despite its fascinating prospects in quantum information processing, optical computing and advanced photonic devices[2-5], a key challenge in harnessing superfluorescence is alleviating its reliance on cryogenic conditions[5-10]. Recently, room-temperature superfluorescence has been successfully achieved using upconverted nanoparticles[11] and quasi two-dimensional lead halide perovskites[12]. These approaches, however, are restricted to a few specific material designs and unsuitable for wide promotion. Here, we report a universal strategy to elevate the operating temperature of superfluorescence. We reveal that the symmetry-protected optical bound state in the continuum (BIC)[13-17] can**

**break the size limitation of superfluorescence ($\lambda^3$)[18] and correlate distant but similar emitters without violating the selection rules, significantly accelerating synchronization process and promoting the possibility of room-temperature superfluorescence. This effect has been experimentally verified using a series of BIC metasurfaces made of different lead halide perovskites. Key features such as the quadratic increase in transient peak intensity and the reduction in pulse width and build-up time at the BIC wavelength confirm the realization of room-temperature superfluorescence that is absent in the pristine material. A theoretical model is also built to explain the experimental observations. This research demonstrates that the operating temperatures of coherent macroscopic states can be effectively improved by artificial field, paving a critical step towards constructing building blocks for optical and quantum applications.**

Creating and manipulating coherent macroscopic state in solids is crucial for emerging photonic and quantum applications.[2-5] It is closely related to a symmetry-breaking phase transition and leads to a number of exceptional collective quantum phenomena, e.g., superconductivity and Bose-Einstein condensation.[19] Superfluorescence is another remarkable example.[1] An incoherent population of dipoles is generated when the system is optically excited. After a delay time or build-up time ($\tau_D \propto \ln(N)/N$, where $N$ is the number of correlated dipoles), phases of dipoles within a tiny volume of $\lambda^3$ can be spontaneously synchronized through vacuum field interactions, forming an ensemble of coherent states with dipole strengths several orders of magnitude higher. Then the system transits into the Dicke superradiant state and produces a short but intense burst of photons.[18] This process is termed as superfluorescence after 1975.[1] According to the Dicke model[18], the peak intensity of the superfluorescence pulse scales as $N^2$ and its radiative pulse width is proportional to $\tau_R \propto 1/N$. Meanwhile, strong light-matter interactions might also introduce a Rabi-type oscillation to the time domain of superfluorescence, first studied by Burnham and Chiao.[20, 21]

In principle, the superfluorescence phase transition is determined by both the establishment of coherence and the dephasing. Superfluorescence will only form when the build-up time $\tau_D$ is shorter than the exciton coherence time ($T_2^*$). Due to the numerous scattering events, the latter one is extremely short in condensed matter, strongly restricting superfluorescence to gas-phase systems and a few solids at cryogenic temperatures.[5-10, 22-25] Operating temperature has been an enduring challenge for practical applications of superfluorescence as solid-state light sources and thus was intensively explored. Over the past few years, great efforts have been achieved in this research area and the operating temperature of superfluorescence is significantly improved.[11, 12, 23] In 2022, Bikiroglu et al. demonstrated room-temperature (300 K) superfluorescence from quasi-two-dimensional hybrid perovskites by binding electronic dipoles to special lattice distortion to isolate the coherent macroscopic state from vibration.[12] By exploiting the unique 4f orbital, which is shielded by the external occupied 5s and 5p orbitals, Haung et al. also succeeded in demonstrating upconversion superfluorescence at 290 K.[11] So far, all current breakthroughs have solely focused on the ingenious design of material with longer dephasing time. The impact of optical field on the synchronization has never been exploited and even considered before. Here, we employ perovskite metasurface to demonstrate that artificial optical field can induce room-temperature superfluorescence that is absent in the pristine material.

In additional to increasing the dephasing time,[11,12] we propose that competition with dephasing can also be achieved by accelerating the synchronization process. In general, the more similar the emitters are, the shorter $\tau_D$ required to synchronize them. When the perovskite is excited by a femtosecond laser, a high density of emitters is instantaneously generated, some of which should be very similar but widely distributed at different locations ($> \lambda^3$). In common vacuum field, such emitters are simply ignored since their interaction requires a propagation term $e^{\pm i\boldsymbol{k}\cdot\boldsymbol{r}}$ and selection rules prohibit the possibility of collective radiation.[18] As depicted in Fig. 1, the situation becomes completely different when the perovskites are prepared into a metasurface with a symmetry-protected bound state in the continuum (BIC). The interaction between photon and exciton can produce a hybrid state with enhanced $T_2^*$.[26-30] Most importantly, the symmetry of metasurface prevents the excitation of the outgoing radiation channels at BIC wavelength and results in an in-plane propagation vector of $\boldsymbol{k} = 0$, thus eliminating the propagation term $e^{\pm i\boldsymbol{k}\cdot\boldsymbol{r}}$ in Dicke's model. In this scenario, the interaction between distant emitters is realized by their destructive interference through continuous spectrum rather than in-plane propagation, so their long-distance correlation is no longer prohibited. The symmetry-protected BIC then functions as a large common field, allowing distant but similar emitters in the field-enhancing region to correlate with each other without violating the selection rules. For such emitters, a much shorter $\tau_D$ is needed to achieve synchronization, so many common materials with small $T_2^*$ can be used to produce superfluorescence at room temperature. In addition, the enhanced coherence of superfluorescence enables far-field directional output along BIC [15] (Schematic in Fig. 1), distinguishing it from conventional photoluminescence (PL).

The above analysis has been verified experimentally. We choose a 100 nm FA-NMA hybrid perovskites film as the light-emitting material (Extended data Fig. 1).[31] The pristine film is optically excited by a frequency doubled femtosecond laser at room temperature (25 °C).[32, 33] The emitted light is collected by an objective lens (20X, NA = 0.4) and analyzed by a streak camera coupled to a monochromator (Methods and Supplementary Note-1). The results are summarized in Extended data Fig. 2. At low pump fluence, the transient emission from the pristine film is a broad PL peak. Then a peak with full width at half maximum (FHWM) of 10 nm appears on its shoulder when the excitation is above 40.6 μJ/cm². The dependence of transient peak intensity on the pump fluence shows that the power slopes of $l$ = 0.98 and $l$ = 8.35, corresponding to PL and amplified stimulated emission (ASE), respectively, ruling out the possibility of superfluorescence. This conclusion is also consistent with the experimentally measured exciton coherence time ($T_2^* = 4\ ps$, see details in Supplementary Note-2)[34], which is two orders of magnitude smaller than Ref. 12 and too small for superfluorescence.

Then the perovskite film is converted to a metasurface by covering it with a 100 nm polymethyl methacrylate (PMMA) film that is patterned with square-latticed periodic air holes (Methods)[32, 33]. Figure 2A shows the top-view scanning electron microscope (SEM) image of the sample. The corresponding high-resolution SEM images are shown as insets, where a lattice size of $p$ = 340 nm and a duty cycle of 0.5 can be seen. The angle-resolved transmission spectra of the metasurface have been measured with a home-made microscope system (Methods) and displayed in Fig. 2B. Several resonances can be clearly observed. The resonant dip at ~550 nm gets narrower and diminishes when the incident angle approaches to $\theta$ = 0. These results, associated with the numerical simulation (Extended data Fig. 3), confirming that one symmetry-protected BIC has been experimentally generated in the gain spectral range of the FA-NMA perovskite.[15]

The metasurface is optically pumped by a femtosecond laser with a beam radius of $R = 5$ μm (Methods) and the spectra of transient emission under different pump fluences are shown in Fig. 2C. The metasurface also produces a broad PL peak at low pump fluence. With the increase of excitation power, a sharp spike with a linewidth of 0.3 nm appears at the BIC wavelength and quickly dominates the emission at higher pump fluences. The peak intensity of the transient emission at the BIC wavelength is analyzed and plotted as a function of the pump density (dots in Fig. 2D). The log-log plot is separated into three regions with different power slopes. The slopes of the first and third regions are $l = 0.94$, $l = 6.3$, corresponding to conventional PL and lasing, respectively. The region with a slope of $l = 2.05$ between PL and lasing, however, has never been reported before. Such a quadratic increase in transient emission is consistent with the collective radiation and indicates the dependence of emission on $N^2$ rather than $N$.[1,18,35]

To confirm the superfluorescence, we characterized the time evolution of emission at the BIC wavelength (Methods). Figure 2E shows the examples of transient emission intensity with excitation fluences of 9.8 μJ/cm$^2$ (bottom panel) and 16.52 μJ/cm$^2$ (top panel), respectively. The decay rate increases at high fluence and results in a shape of burst. A second peak appears at its shoulder, which has the same pulse width but much weaker intensity, corresponding to the well-known Burnham-Chiao ringing. Following the equation $I(t) = A \operatorname{sech}^2\left(\frac{1}{\tau_R}(t - \tau_D)\right)\left(1 - \frac{t}{T_2^*}\right)$, the temporal pulse width $\tau_R$ and the build-up time $\tau_D$ have been extracted by fitting the curves at different pump fluences.[1,23] All the results are plotted as dots in Fig. 2F. As the excitation intensity increases, the build-up time reduces from ~6.2 $ps$ to ~2 $ps$ (top panel), and the pulse width decreases from ~5.8 $ps$ to ~2.8 $ps$ (bottom panel). The solid lines in Fig. 2F are their fitting curves, which conform to the trends of $\ln(N)/N$ and $1/N$, respectively, and are in verry good agreement with the predictions of superfluorescence.[1,18,35]

In principle, superfluorescence exhibits a longer coherence time than PL. This information has also been experimentally confirmed through back focal plane imaging.[15] The results are shown in Fig. 3A, where the excitation is only 14.7 μJ/cm² (below the laser threshold). A donut-shaped beam profile with a divergent angle of 2° (Panel I) can be seen in the back focal plane. After passing a linear polarizer, the donut changes to two lobes that rotate with the axis of polarizer (Panels II-V), demonstrating a radial polarization. Then the donut beam is injected into an off-axis Michelson interferometer and spatially extended interference fringes can be observed (Panel VI). By adjusting the delay time of one arm of the Michelson interferometer, the interference pattern is recorded and the fringe contrast at each delay is plotted in Fig. 3B. The first-order correlation of the donut is much longer than conventional PL (open squares). Up to now, all the experimental observations prior to laser emission, including quadratic increase of transient emission intensity, Burnham-Chiao ringing, reduced $\tau_R$ and $\tau_D$, polarization state, far-field directionality, and interference pattern, are all consistent with the collective radiation and demonstrate that room-temperature superfluorescence has been successfully realized at the BIC wavelength.

Based on the confirmation of room-temperature superfluorescence, we then gradually decreased the radius of pump area and explored the variation of superfluorescence threshold of the same metasurface. Following previous studies[11, 12, 22, 23], the superfluorescence threshold $P_{th}$ is defined as the transition point from $l$ = 0.94 to $l$ = 2.05 (see accurate definition in Supplementary Note-3). All results are shown as dots in Fig. 4A. As the pump spot decreases, the superfluorescence threshold remains unchanged at first, rises sharply when $R$ is less than 5.75 μm, and finally disappears after $R$ < 3.5 μm, where the power slope reduces from $l$ = 1.5 - 2.0 to $l$ ~ 1.0 (Open squares in Fig. 4A, see an example in Extended data Fig. 5). This is a direct proof that the superfluorescence from our metasurface arises from the synchronization of distant emitters ($> \lambda^3$) via the symmetry-protected BIC, and the size limitation ($< \lambda^3$) of conventional superfluorescence has been broken.

To understand the superfluorescence, the exciton coherence time $T_2^*$ of the perovskite metasurface is measured (Methods and Supplementary Note-2). When the delay is smaller than $T_2^*$, the second pulse can coherently interact with the excitons generated by the first pump, resulting in oscillations in the total intensity of the nonlinear signal. One example is shown in Fig. 4B, which gives an exciton coherent time of $T_2^* \sim 15\,ps$. This value is larger than $\tau_D$ and $\tau_R$ in Fig. 2 and consistent with the fitted $T_2^*$ (Supplementary Note-2) from the time-resolved curve. The enhanced $T_2^*$ can be attributed to exciton-photon interactions in metasurfaces (see Supplementary Note-4).[26-30] But the value is still more than an order of magnitude smaller than Ref. 12 and insufficient to support superfluorescence at room temperature. With the experimental parameters, we started with Dicke's type interaction Hamiltonian and built a theoretical model for the superfluorescence.[1,18] As stated above, the symmetry-protected BIC enables the interaction between distant emitters via the external continuum and breaks the size restriction of $\lambda^3$. In this sense, the synchronization shall occur within similar emitters rather than the completely random ones. Then the master equations are the same as Ref. 1 except that the random phase of emitters is further restricted by a factor $\alpha$ (Supplementary Note-5). With the increase of emitter similarity (reduction of randomness), the required $T_2^*$ for a burst keeps decreasing and can be smaller than the value of our metasurface ($T_2^* \sim 15\,ps$), enabling superfluorescence at room temperature (Fig. 4C, Supplementary Note-5).

    In a large area, many emitters can find their partners similar to themselves. Such a probability increases as the density of emitters is higher. Basically, the more emitters participate in superfluorescence, the fewer emitters participate in conventional PL. In this sense, we can even expect the reduction of PL intensity when superfluorescence occurs in our BIC metasurface. This inference has been experimentally confirmed. The open squares in Fig. 2D show the integration of transient emission intensity in a spectral range from 530 nm to 540nm. An obvious reduction of PL intensity can be seen when the quadratic increase ($l \sim 2$) emerges. This is in stark contrast to the PL of the pristine film (Extended Data Fig. 2) and the grating mode off the Γ point (Extended Data Fig. 4). Using a similar method, we also investigated the dependence of the power slope of PL before laser emission on the pump beam size (Fig. 4D). The appearance of superfluorescence at BIC wavelength is always associated with a negative power slope of PL region (reduction in intensity). When $R < 3.5$ μm and superfluorescence disappears, the transient fluorescence intensity also increases monotonically without reduction (details in Supplementary Note-6).

    In our metasurface, the size restriction is broken only at the BIC wavelength, so the

wavelength of superfluorescence can be programmably tuned via structural parameters. This is completely different from the conventional superfluorescence. In experiment, we realized this function by fabricating a series of metasurfaces with different lattice sizes (Supplementary Note-7). As the lattice size increase from $p$ = 333 nm to $p$ = 355 nm, the emission wavelength of superfluorescence linearly shifts from 543 nm to 556 nm (Fig. 5A). The studies on the power-dependent transient emission intensity before lasing action ($l \gg 1$) show that the power slope remains around $l$ = 1.5 to $l$ = 2.0 (dots in Fig. 5B), confirming the room-temperature superfluorescence in such metasurfaces well. Then we know that the BIC induced superfluorescence is generic and its wavelength can be precisely tuned in a broad spectral range.

For the cases of $p \leq 325\ nm$ and $p \geq 360\ nm$, the power slope drops to $l \sim 1$ and superfluorescence disappears. The vanishment of superfluorescence in these metasurfaces is caused by the reduction of exciton coherence time $T_2^*$. The change of lattice size affects the resonant wavelength and thus changes $T_2^*$ of the hybrid state of exciton-photon interaction. As open squares shown in Fig. 5B, $T_2^*$ remains around 9 – 15 $ps$ and drops to below 8.7 $ps$ at $p$ < 330 nm to $p$ > 355 nm, where the superfluorescence disappears. The increase of temperature will decrease the $T_2^*$ of pristine materials and thus affects effective $T_2^*$ of the perovskite metasurface too. The dots in Fig. 5C shows the power slope slightly below lasing action at different temperature. The power slope remains around $l$ = 1.5 – 2.0 at first and then drops to $l \sim 1$ at 45 °C. The corresponding $T_2^*$, which is measured and plotted as open squares in Fig. 5C, also gives an exciton coherence time of $T_2^*$ = 8.7 $ps$ at 45 °C. Then we know that our metasurface also has a minimal requirement of $T_2^* \sim 8.7$ ps, which can be improved by increasing the Q factor of symmetry-protected BIC.

In summary, we have demonstrated a novel and universal mechanism to generate coherent macroscopic state at room or even higher temperature. Compared to ingenious material design, symmetry-protected BIC can overcome the size limitations of traditional superfluorescence and significantly accelerate the establishment of synchronization. Simultaneously, exciton-photon interaction also enhances the exciton coherence time $T_2^*$. The combination of two effects enables perovskite-based BIC metasurfaces to produce room-temperature superfluorescence not present in their original materials. This new mechanism is also robust to different perovskites (Extended data Figs. 5-9). It can be promoted to other quantum systems[18] as well and boost the advancements of emerging quantum applications[2-5, 36].

**Methods:**

**Numerical simulation**. The band structures and quality (Q) factors of the BIC metasurface are calculated with a finite-element method based commercial software (COMSOL Multiphysics). Periodic boundary condition is applied in *x*- and *y*- directions to mimic the infinitely large sample sizes. Perfectly matched layers are employed in z direction to absorb the outgoing waves. The optical constants of perovskite films were obtained from ellipsometry measurement. The refractive indices of glass substrate and polymer grating are fixed at n = 1.45 and n =1.54, respectively. The environment is air with n = 1. The eigen-frequency solver in COMSOL produces complex valued eigenfrequency ($\omega$). Then the resonant wavelength and Q factor can be achieved from $2\pi c/Re(\omega)$ and $Q = Re(\omega)/|2Im(\omega)|$, respectively. With the gradual change on incident angle, the band structure and the corresponding Q factors are obtained.

**Fabrication of perovskite metasurface**. The BIC metasurface is fabricated with a etchless process. Basically, the precursor solutions of N2F8 ((NMA)2FAn-1PbBr3n+1, n=8) perovskite was achieved by adding a 25% molar ratio of 1-naphthylmethylamine bromide (NMABr) into a 1:1 ratio of HC(NH2)2Br (FABr) and PbBr2 in DMF at 0.4 M and stirred at 60 °C for 12 h. The perovskite film was prepared using a one-step spin-coating at 5,000 r.p.m. for 30 s. During the spin coating, 0.3 ml of ethyl acetate is dropped onto the perovskite precursor layer. The substrate is baked on a hotplate at 85 °C for 15 min and then the perovskite film is obtained. The thickness of the perovskite film is 100 nm. Then 100 nm electron-beam resist (ZEP260A) is spin-coated onto the perovskite film and placed at room temperature overnight. The electron-beam resist is patterned with electron-beam aligner (Raith e-LINE, 30 kV). After developing in N50 for 60 s and dried with nitrogen gas, the exposed area is removed and the BIC metasurface can be finally obtained.

**Optical characterization**. In optical experiments, the sample is mounted on a rotational translation stage and measured with a home-made microscope. The white light source (Thorlabs SLS201L/M) is focused by a 20X objective lens onto the sample at a specific angle. The transmission is collected by an optical lens and recorded by a CCD (PIXIS 256, Princeton Instruments) coupled to a spectrometer (ISO Plane SCT320, Princeton Instruments). Then the sample is rotated to achieve the angle-dependent transmission spectra.

For the characterization of superfluorescence, the Ti:Sapphire laser (800 nm, 100 fs pulse width, 1 kHz repetition rate) is frequency doubled after passing a BBO nonlinear crystals. The laser beam is focused by a 20X objective lens onto the sample. A white light source is added to the optical setup via a 50:50 beam splitter for accurately determining the pump position. The emission from the sample is collected and collimated by the second 20X objective lens. After passing a long-pass filter, the back focal plane image is achieved by a 4-*f* system and a CMOS camera. The donut in the back focal plane can be spatially selected and recorded by a streak camera (XIOPM 5200) coupled to a spectrometer (Horiba iHR320). All the experiments are carried under ambient conditions (with a temperature of 300 K and a humidity of 45%).

The exciton coherence time $T_2^*$ is measured with a variation on standard pump-probe measurements. Two femtosecond laser pulses at 1054 nm with the same intensity and a delay time ΔT excite the sample. Due to the resonant excitation, each pulse can individually produce an intense and coherent nonlinear signals at ~ 530 nm and a broad band PL background. The generated nonlinear signals accounts for about 63% of the total energy and its intensity is defined as $I_0$. In the pump-probe experiment, when $\Delta T \leq T_2^*$, the excitons generated by the first pump pulse can interact with the second pump and affects the intensity of the generated nonlinear signals. In this sense, the intensity of total coherent pulses should oscillate within the exciton coherence time, and thus $T_2^*$ can be obtained by fitting the envelope of the oscillations.

## Data availability

The data that support the plots within this paper and other finding of this study are available from the corresponding author upon reasonable request.

## Acknowledgements

This research was supported by National Key Research and Development Program of China (grant no. 2024YFB2809200), National Natural Science Foundation of China (grant nos. 12025402, 12334016, 62125501, 62305084, 6233000076, 12261131500, 92250302 and 11934012), New Cornerstone Science Foundation through XPLORER PRIZE, Shenzhen Fundamental Research Project (grant nos. JCYJ20241202123729038, JCYJ20241202123719025, JCYJ20220818102218040 and GXWD20220817145518001), Fundamental Research Funds for the Central Universities (grant no. 2022FRRK030004), and Guangdong Basic and Applied Basic Research Foundation (2023A1515011746).

## Author contributions

Q.S. conceived the idea and supervised the project. H.T., C.H. Q.S. designed the experiments. H.T. conducted the simulations. H.B. Q.S. built the theoretical model. H.T. and X.J. performed the experiments. H.T., C.H., S.X., S.Y., J.H. Q.S. analyzed the data. Q.S. drafted the paper with inputs from all authors.

## Competing interests

The authors declare no competing interests.

**Figure captions:**

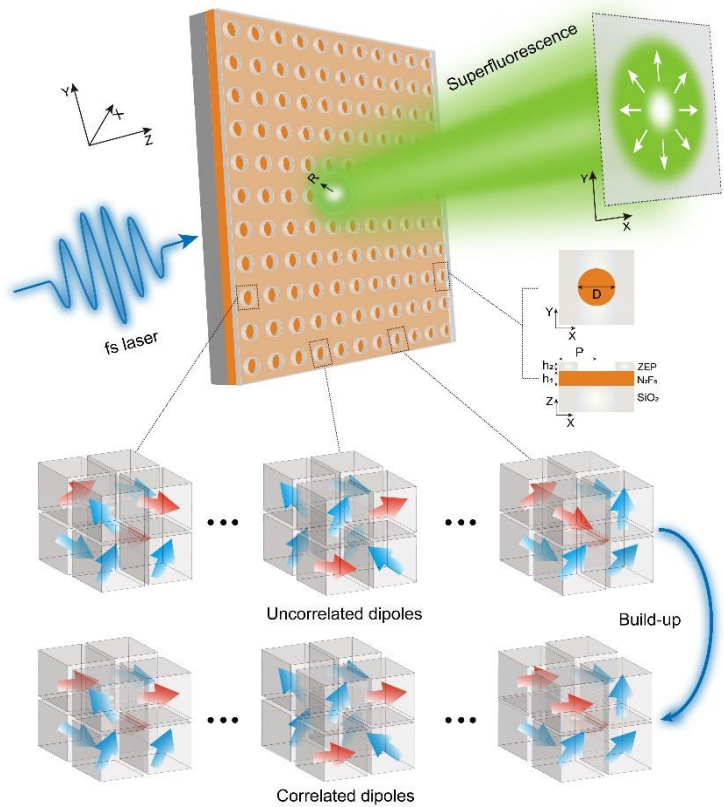

**Figure 1. Schematic of symmetry-protected BIC induced superfluorescence.** Optical excitation (with a radius of *R*) produces a huge number of uncorrelated emitters in the perovskite membrane, and some of which are quite similar but spatially far apart (red arrows). Their emission with non-zero in-plane propagation vector is inhibited by the band structure, whereas the vertical radiation is coherently linked by the continuum spectrum, thus breaking the size limitation of conventional superfluorescence ($< \lambda^3$). Here, symmetry-protected BIC acts as an artificial common field, connecting and correlating these similar emitters without violating selection rules. The more similar the emitter are, the shorter time it takes to synchronize them. The emitters marked by red arrows can thus be synchronized in a time ($\tau_D$) much shorter than the exciton coherence time ($T_2^*$), resulting in room-temperature superfluorescence not present in the original material.

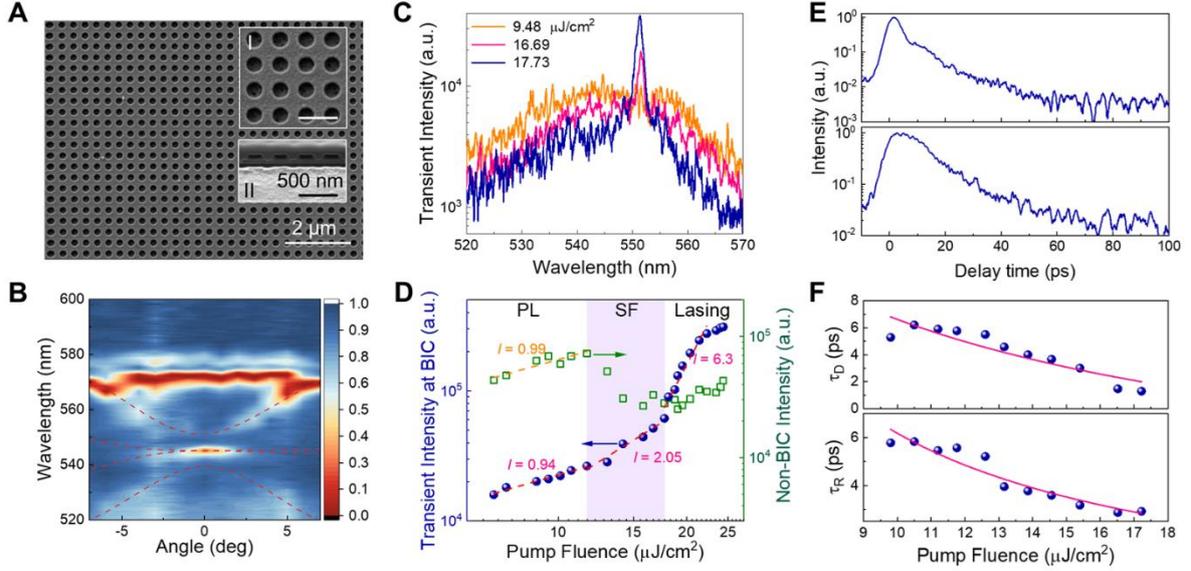

**Figure 2. Optical characterization of a perovskite-based BIC metasurface. A**. Top-view SEM image of top nanostructure of the BIC metasurface. High-resolution top-view and side-view SEM images are shown as insets. **B**. Optically recorded angle-resolved spectra of the BIC metasurface. The numerically calculated band structures are overlayed as dashed lines. **C**. The transient emission spectra from the metasurface under different pump fluences. **D**. The dependence of peak intensity of transient emission at the BIC wavelength (dots) and non-resonant spectral range from 530 nm to 540 nm (open squares) on the pump fluence. The transition point from $l = 0.94$ to $l = 2.05$ is defined as superfluorescence threshold $P_{th}$. **E**. Transient emission intensity at the BIC wavelength under different pump fluences of 9.8 μJ/cm² (bottom panel) and 16.52 μJ/cm² (top panel). **F**. The extracted build-up time $\tau_D$ (top panel) and $\tau_R$ as a function of pump fluence. The solid lines correspond to the fitted curves of $\ln(N)/N$ and $1/N$, respectively.

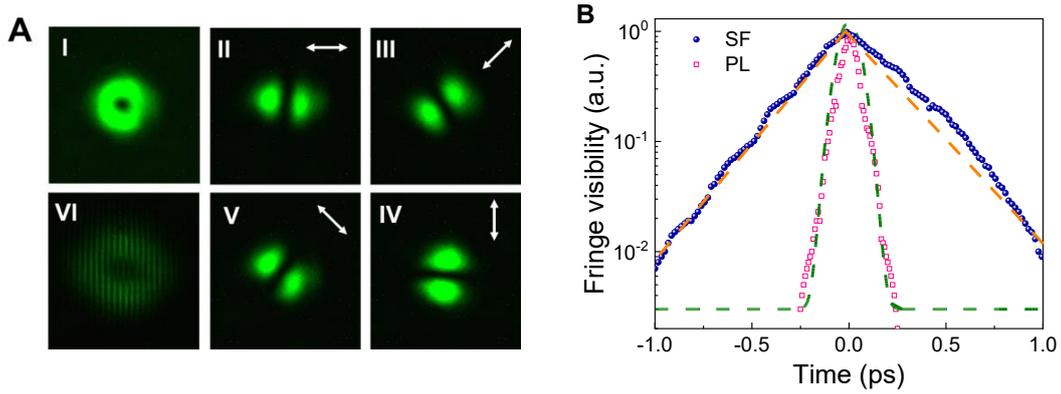

**Figure 3. Further confirmation of room-temperature superfluorescence. A**. Back focal plane image of emission at BIC wavelength (Panel I) and the images passing through a linear polarizer with axis along 0º (Panel II), 45º (Panel III), 90º (Panel IV), and 135º (Panel V). Panel VI is the self-interference pattern of the donut. Here the excitation is only 14.7 μJ/cm$^2$, which is below the laser threshold. **B**. First-order correlation of the emission from BIC mode (dots) and a pristine film (open squares).

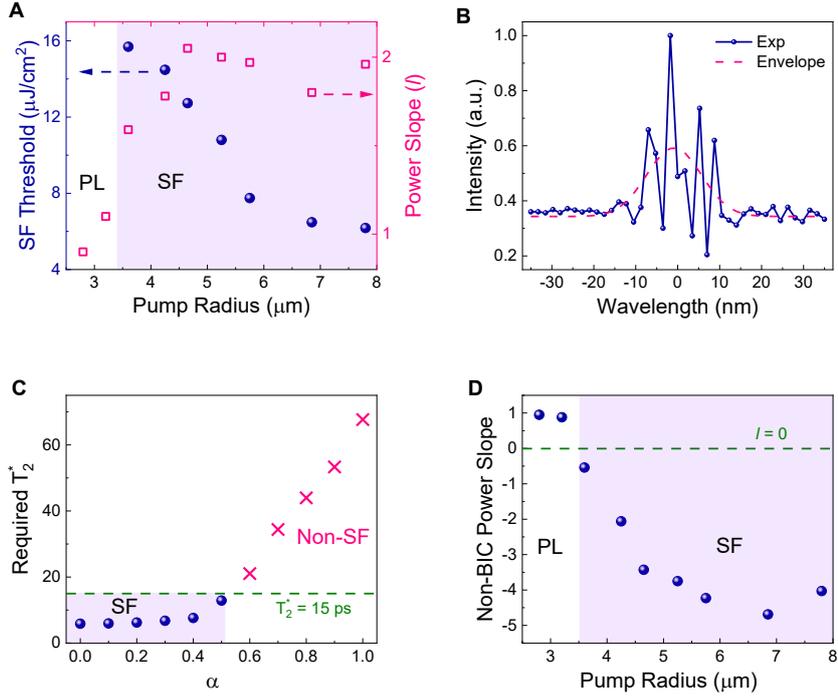

**Figure 4. Mechanism for room-temperature superfluorescence.** **A**. Dependence of superfluorescence threshold $P_{th}$ (dots) and power slope (open squares) on the size of pump area $R$. **B**. The dependence of nonlinear signal intensity on the delay between two pumps. The exciton coherence time of $T_2^*$ can be obtained by fitting the envelope of the oscillations (red dashed line). **C**. Required exciton coherence time $T_2^*$ for superfluorescence as a function of randomness of the emitters. The dashed line corresponds to the measured exciton coherence time of our perovskite metasurface. **D**. The power slope of emission at non-resonant wavelength as a function of pump size. Here the excitation is around 0.9 times of the lasing threshold. The superfluorescence region is highlighted by a violet shadow.

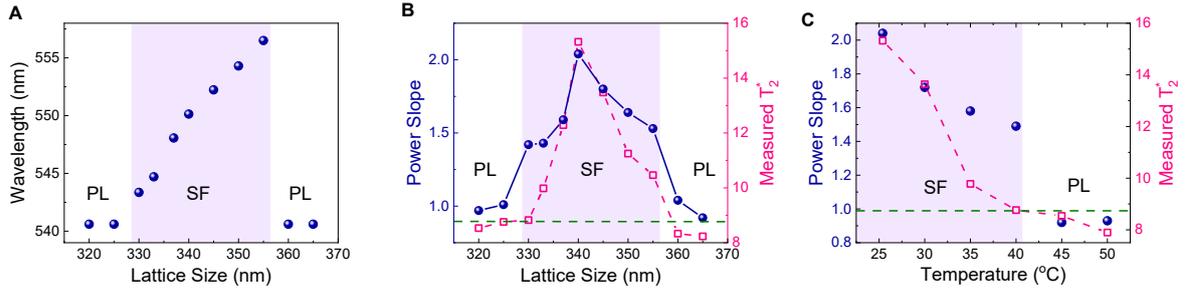

**Figure 5. Superfluorescence of different metasurfaces. A**. The peak wavelength of transient emission of the metasurface a function of lattice size *p*. Here the excitation is below the laser threshold. PL has a fixed wavelength, whereas the superfluorescence increases linearly with the lattice size. **B**. The experimentally recorded power slope of transient emission (dots) before laser emission in metasurfaces with different lattice sizes. **C**. The experimentally recorded power slope of transient emission before lasing (dots) of one metasurface at different temperature. Open squares in B and C are the corresponding dephasing time of such metasurface. The green lines correspond to $T_2^* = 8.7\,ps$. The superfluorescence region is highlighted by a violet shadow.

**Extended data**

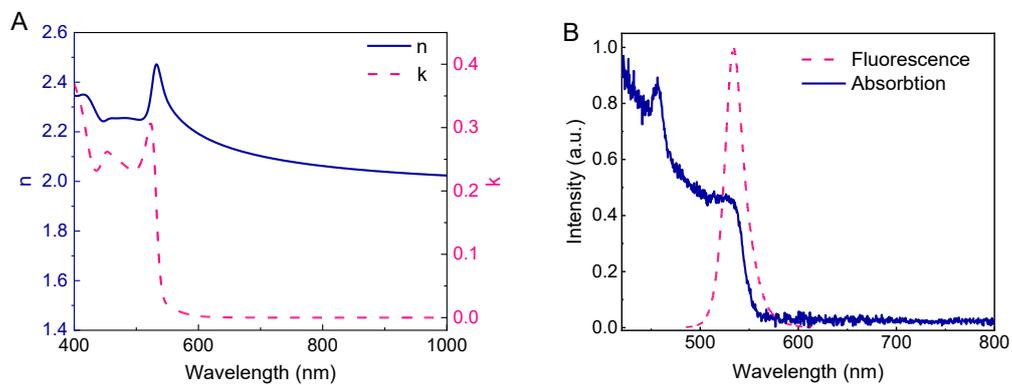

**Extended data Fig. 1. Optical properties of FA-NMA hybrid perovskite. A**. The refractive index and light extinction coefficient of the hybrid perovskite film. **B**. The absorption and photoluminescence of the hybrid perovskite film. **Since the wavelengths of fluorescence and band edge are close to bulk perovskite, there is no obvious quantum confinement effect.**

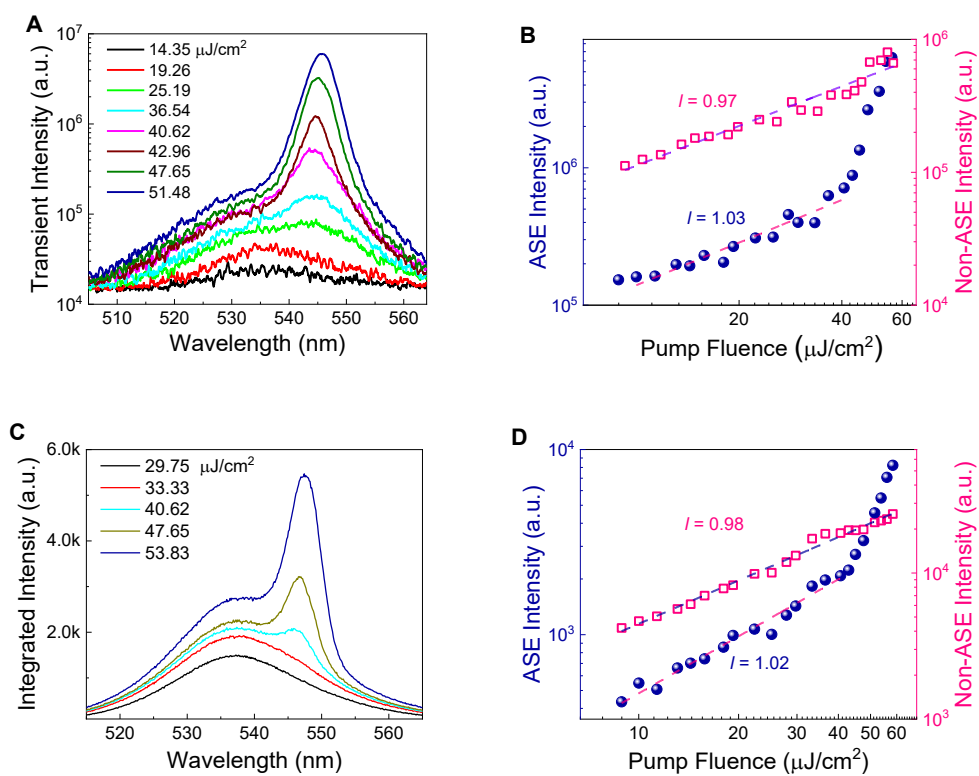

**Extended data Fig. 2. Time-resolved measurement of a bare FA-NMA hybrid perovskite film. A**. Transient emission spectra from the bare perovskite film under different pump fluences. **B**. The dependence of transient emission at the ASE wavelength on the pump fluences. Open squares correspond to the intensity of transient emission within the spectral range from 520 nm to 530 nm as a function of pump fluence. Different from the superfluorescence in the main text, the intensity increases monotonically and no reduction can be observed. **C** and **D** are the corresponding time-integrated and the threshold curves at ASE wavelength and PL wavelength. **Below the ASE threshold, the power slope of transient emission and integrated emission are almost the same. This shows that FA-NMA perovskite cannot support either superfluorescence or bimolecular emission.**

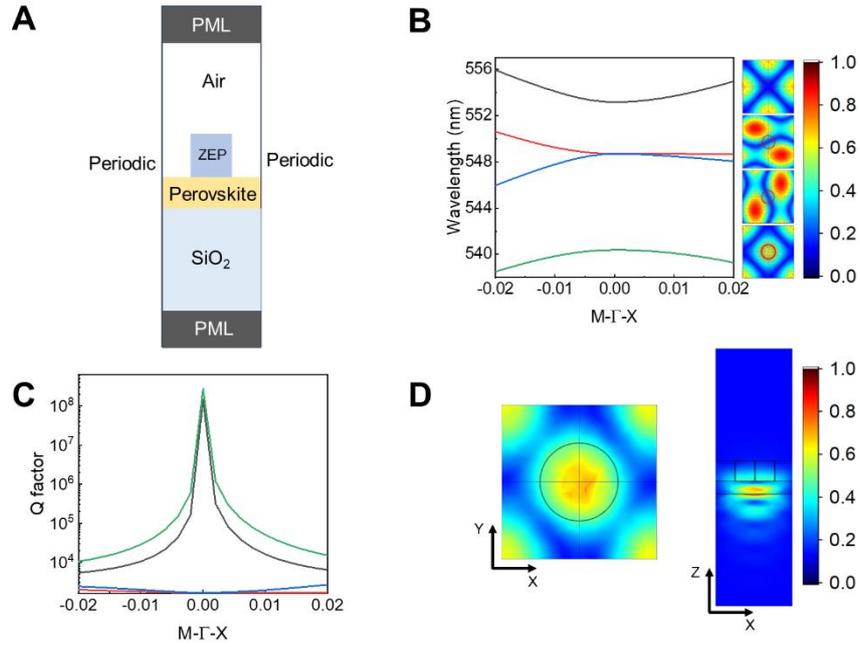

**Extended data Fig. 3. Numerical design of BIC metasurface. A**. Schematic of one unit cell in the perovskite metasurface. **B**. Numerically calculated band structure of the metasurface. **C**. The Q factors of the resonances in B. Ultrahigh Q values have been obtained from two of them, indicating the realization of BIC mode. **D**. The field distributions of the BIC in *x-y* (left panel) and *xz* (right panel) plane, respectively. **It is obviously to see that the power of resonant mode is mostly confined within the gain layer, *i.e.*, perovskite film.**

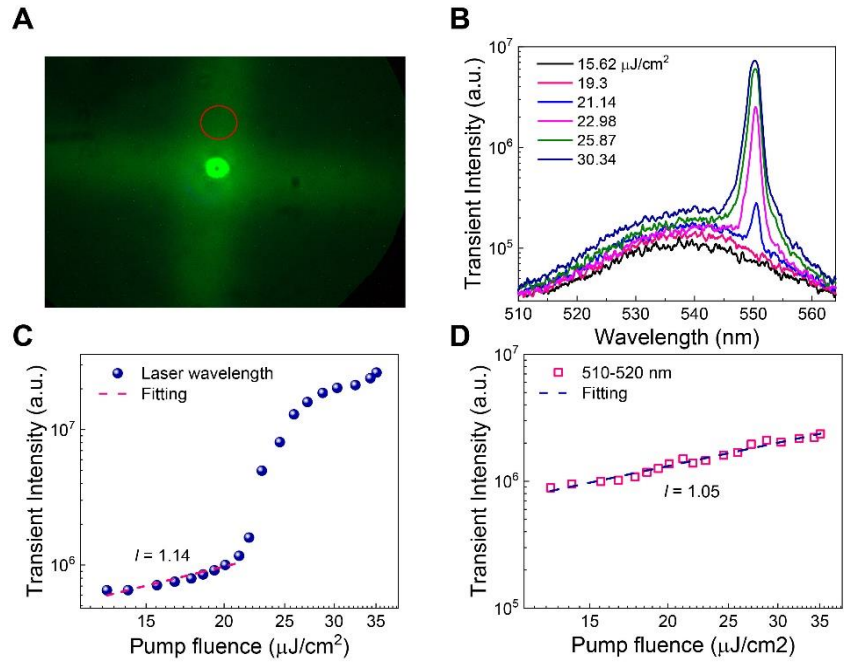

**Extended Data Fig. 4. Control experiment at the grating mode in the same metasurface.** **A**. The back focal plane image of the emission. The dashed circle marks the detection area where the grating mode radiates. **B**. The transient emission spectrum recorded at the dashed circles. **C** and **D** are dependence of transient peak intensity at the grating mode laser and the integrated emission from 510 nm to 520 nm on the pump fluence. No quadratic increase with a power slope of $l = 2$ can be seen in **C**. Meanwhile, the intensity of PL (open squares in **D**) increases monotonically. Both are completely different from the results in Fig. 2.

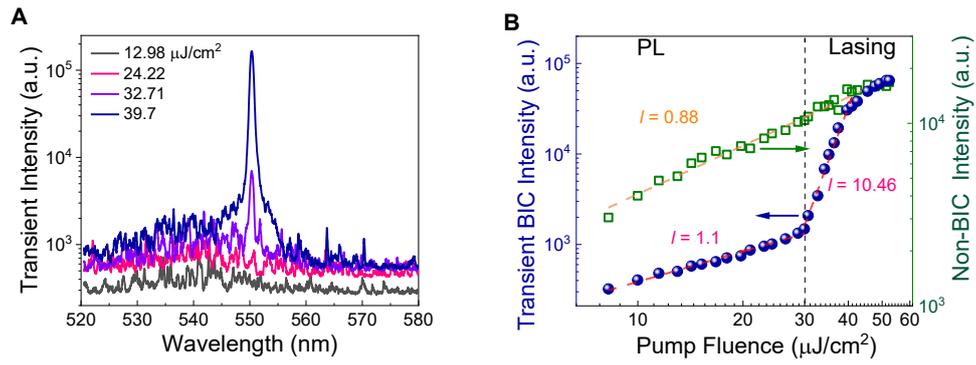

**Extended data Fig. 5. Emission from the same metasurface with a pump radius of *R* = 3.2 μm. A**. Transient emission spectra at different pump fluences. **B**. The dependence of transient emission peak intensity at resonance wavelength (dots) and at 530 nm-540 nm (open squares) on the pump fluence.

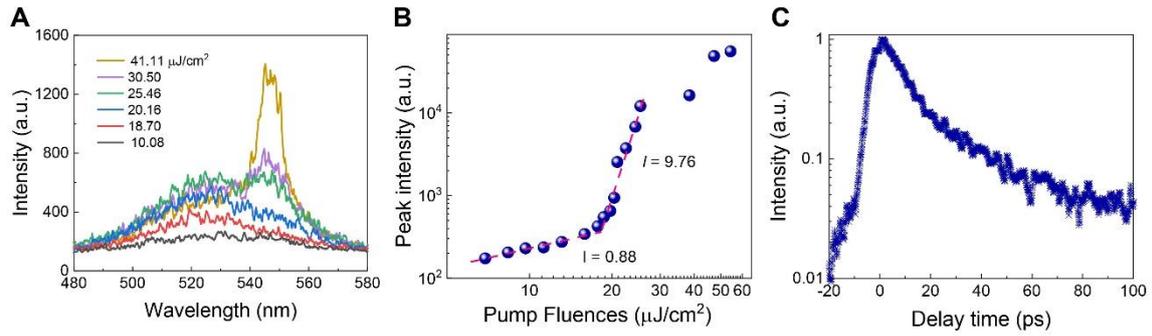

**Extended data Fig. 6. Control experiment on time-resolved measurement of a bare MAPbBr$_3$ perovskite film at room temperature. A**. The emission spectra from the bare perovskite film under different pump fluences. **B**. The peak intensity of transient emission as a function of pump fluence. **C**. The time-resolved emission spectra at ASE wavelength. **Different from BIC metasurface in the main text and below, no Burnham-Chiao ringing has been observed here and the extracted slope below the ASE threshold is close to 1.**

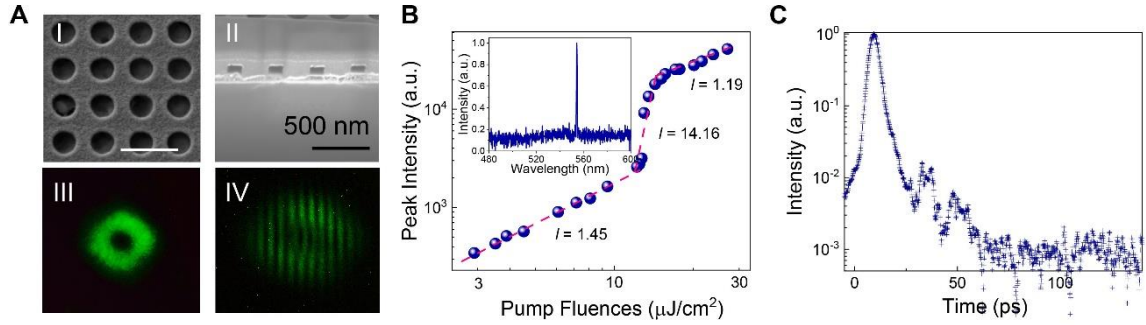

**Extended data Fig. 7. Room-temperature superfluorescence from a MAPbBr$_3$ BIC metasurface. A**. Panels-I and II are the high-resolution SEM images of a new metasurface. Panels-III and IV show the back-focal plane image and the corresponding self-interference pattern when the excitation is 9 μJ/cm$^2$. **B**. The peak intensity of transient emission as a function of pump fluence. Inset shows the emission spectrum of the donut at 9 μJ/cm$^2$. Note that only the emission at the donut region is analyzed. The incoherent PL is negligibly small and only the coherent superfluorescence is plotted here. Detailed information can be seen in Supplementary Note-3. **C**. Transient intensity at the BIC wavelength under a pump fluences of 9 μJ/cm$^2$. Here the lattice size is $p$ = 325 nm and the duty cycle is 0.5. **Both of the power slope of $l$ = 1.45 and the burst in the time-resolved spectrum confirm the existence of superfluorescence in our MaPbBr$_3$ metasurface. This result shows that our mechanism is quite generic and can be applied to different materials.**

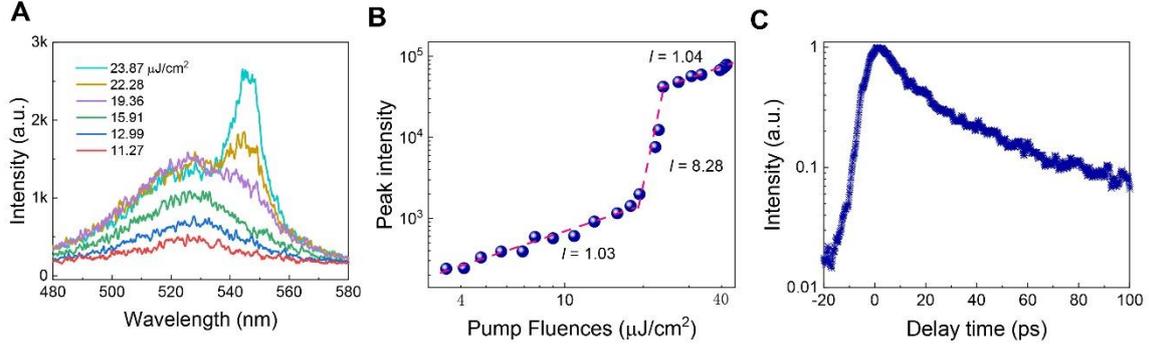

**Extended data Fig. 8. Control experiment on time-resolved measurement of a bare FAPbBr$_3$ perovskite film at room temperature.** **A**. The emission spectra from the bare perovskite film under different pump fluences. **B**. The peak intensity of transient emission as a function of pump fluence. **C**. The time-resolved emission spectra at ASE wavelength. **Different from BIC metasurface in the main text and below, no Burnham-Chiao ringing has been observed here and the extracted slope below the ASE threshold is close to 1.** Similar to bare MAPbBr$_3$ film, several discrete peaks can be seen at high pump fluences. Associated with the "S"-shaped curve in log-log plot (B), we know that random lasers have been generated in FAPbBr$_3$ film too. However, no superfluorescence can be obtained from the FAPbBr$_3$ film.

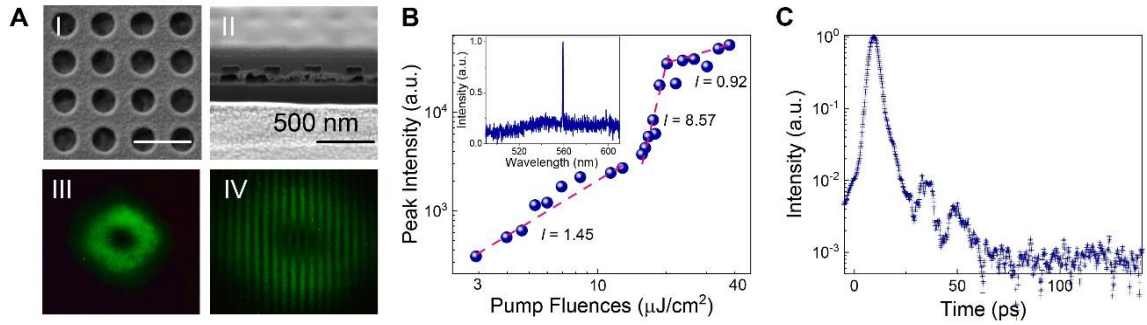

**Extended data Fig. 9. Room-temperature superfluorescence from a FAPbBr$_3$ BIC metasurface. A**. Panels-I and II are the high-resolution SEM images of a new metasurface. Panels-III and IV show the back-focal plane image and the corresponding self-interference pattern when the excitation is 12 μJ/cm$^2$. **B**. The peak intensity of transient emission as a function of pump fluence. Inset shows the emission spectrum of the donut at 12 μJ/cm$^2$. **C**. Transient intensity at the BIC wavelength under a pump fluences of 12 μJ/cm$^2$. Here the lattice size is $p$ = 330 nm and the duty cycle is 0.5. **Similar to MAPbBr$_3$ in Extended data Fig. 7, only the emission of donut is analyzed and plotted. We can see a power slope of $l$ = 1.45 and a superfluorescent burst obtained from the FAPbBr$_3$ BIC metasurface.** Once again, these observations and the results of ASE in bare film confirm that the symmetry-protected BIC ca break the size limitation of conventional superfluorescence and correlate distant (but similar) emitters in a much shorter time, resulting in superfluorescence that is absent in its original materials.

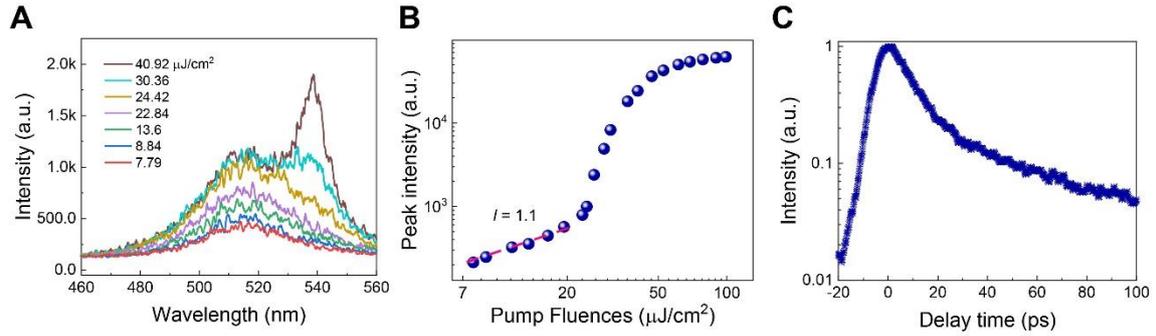

**Extended data Fig. 10. Control experiment on time-resolved measurement of a bare CsPbBr₃ perovskite film at room temperature.** **A**. The emission spectra from the bare perovskite film under different pump fluences. **B**. The peak intensity of transient emission as a function of pump fluence. **C**. The time-resolved emission spectra at ASE wavelength. **Different from BIC metasurface, no Burnham-Chiao ringing has been observed here and the extracted slope below the ASE threshold is close to 1.**

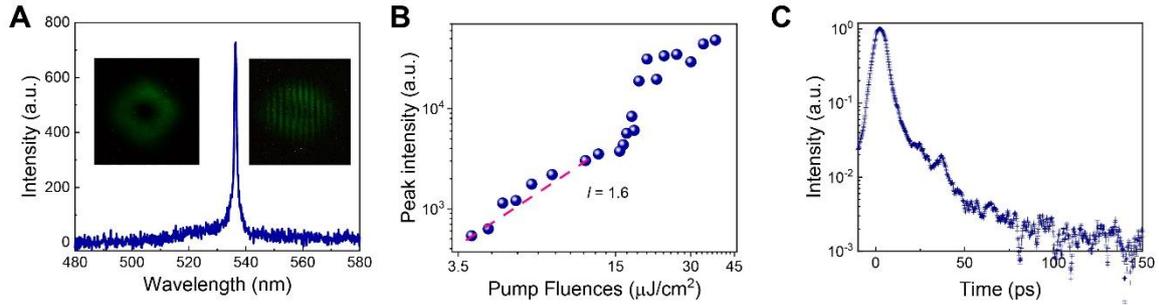

**Extended data Fig. 11. Room-temperature superfluorescence from an all-inorganic perovskite CsPbBr$_3$ based BIC metasurface. A**. The emission spectrum of the metasurface at a pump fluence of 14 μJ/cm$^2$. Insets show the back-focal plane image and the corresponding self-interference pattern when the excitation is 14 μJ/cm$^2$. **B**. The peak intensity of transient emission as a function of pump fluence. A below laser threshold of 1.6 can be clearly seen. **C**. Transient intensity at the BIC wavelength under a pump fluences of 14 μJ/cm$^2$. Here only the emission of the donut is analyzed and plotted. **As a result, the power slope of *l* = 1.6 (before lasing) and the burst in the time-resolved spectrum confirm the existence of room-temperature superfluorescence in CsPbBr$_3$ BIC metasurface.**